\documentclass[aps,prd,reprint,10pt,amsmath,amssymb,superscriptaddress]{revtex4-2}
\usepackage[utf8]{inputenc}
\usepackage[T1]{fontenc}
\usepackage[english]{babel}
\usepackage{amsmath,amssymb,amsfonts}
\usepackage{physics}
\usepackage{mathrsfs}
\usepackage{graphicx}
\usepackage{xcolor}
\usepackage[hidelinks]{hyperref}
\newcommand{\lmatt}{\mathcal{L}_\text{matt}}
\newcommand{\lmattstar}{\mathcal{L}_{\text{matt}^{*}}}
\begin{document}
\title{Weak-field analytical solutions for $f(R,\lmatt)$ gravity in a static and spherically symmetric space-time for perfect fluids}

\author{Sergio Mendoza}
\email{sergio@astro.unam.mx}
\affiliation{Instituto de Astronom\'{\i}a, Universidad Nacional Aut\'onoma de M\'exico, Ciudad Universitaria 04510, M\'exico}

\author{Roberto Urrutia}
\email{rurrutia@astro.unam.mx}
\affiliation{Instituto de Astronom\'{\i}a, Universidad Nacional Aut\'onoma de M\'exico, Ciudad Universitaria 04510, M\'exico}

\begin{abstract}

We study the weak-field regime of $f(R,\lmatt)$ gravity for a static and spherically symmetric configuration described by a perfect fluid, whose energy density is perturbed about a constant reference value. By means of a 
weak-field limit expansion about a Minkowski geometric background and under a quasi-Newtonian approximation, we derive and analytically solve the coupled system that governs the gravitational perturbations and the matter variables.

The solution contains the long-range Newtonian contribution and Yukawa-type corrections in the temporal and radial components of the metric. In the matter sector, the energy density, the internal energy, and the pressure acquire radial contributions, with a non-zero constant background rest-mass energy density and vanishing background internal energy and pressure. The results show how the non-minimal coupling between curvature and matter jointly modifies the geometry and the fluid in the considered regime.

\par\vspace{1em}

\end{abstract}

\maketitle

\section{Introduction}

Black holes constitute one of the fundamental predictions
of general relativity. In the static and spherically symmetric case, the Schwarzschild solution describes the exterior gravitational field of an isolated compact object in 
vacuum~\cite{Schwarzschild1916} and constitutes the natural reference point to study possible deviations from the predictions of general relativity.

In astrophysical scenarios, compact objects can be found
immersed in different matter distributions, such as gas, accretion
flows, or stellar populations. These components can contribute
to the gravitational field and modify the surrounding geometry. Therefore, it is
interesting to study configurations in which the geometry and the matter sector are treated jointly.

This problem acquires special interest in extended theories of gravity that incorporate a direct coupling between geometry, through curvature, and matter. Among them is $f(R,\lmatt)$ gravity, whose action depends on a general function of the Ricci scalar $R$ and the matter Lagrangian $\lmatt$ ~\cite{HarkoLobo2010}. This dependence introduces a non-minimal coupling between both sectors and modifies both the field equations and the effective dynamics of matter~\cite{HarkoLobo2010,BertolamiHarkoLoboParamos2008}.

In the weak-field regime, different modified theories of gravity can  introduce Yukawa-like corrections associated with additional gravitational degrees of freedom~\cite{Stelle1978}. In particular, for a general class of analytic metric $f(R)$ theories, the gravitational potential can contain exponentially decreasing contributions superimposed on the Newtonian term~\cite{CapozzielloStabileTroisi2009}. These contributions introduce
characteristic length scales and allow different 
finite-range modifications from the long-range gravitational interaction.

In this work we study whether a similar structure appears in
$f(R,\lmatt)$ gravity for a static and spherically
symmetric configuration described by a perfect fluid, whose energy density is perturbed about a constant reference value. To this end,
we develop a post-Newtonian (PN) expansion about a Minkowski geometric background and, under a quasi-Newtonian (QN) approximation, we derive and solve the coupled system that governs the gravitational perturbations and the fluid variables. The resulting solution contains the long-range Newtonian contribution and Yukawa-like corrections in the metric, accompanied by radial contributions in the energy density, the internal energy, and the pressure for a perfect fluid.

The paper is organised as follows. In Section II we formulate the weak-field limit of $f(R,\lmatt)$ gravity. Starting from the action and the associated field equations, we introduce the post-Newtonian expansion about a Minkowski background for a static, spherically symmetric perfect fluid, organise the field equations order by order, and derive the corresponding linearised equations. We then separate the constant background from the radial perturbations in both the geometric and matter sectors, and we establish the quasi-Newtonian regime by controlling the normalised residuals of the reduced metric equations on a compact radial interval. In Section III we solve the resulting reduced system analytically. We obtain the general solution, derive sufficient weak-field bounds, and select the asymptotically flat branch; in this branch the constant matter background is dust-like, with non-vanishing rest-mass energy density and vanishing internal energy density and pressure. We also identify a particular branch satisfying a $\kappa$-law relation between pressure and internal energy density and show that it admits a local first-order polytropic approximation at a finite reference radius. In Section IV we construct the corresponding weak-field line element and the extended Newtonian potential, and we discuss the conditions under which the Yukawa-type correction is attractive or repulsive. Finally, Section V presents our conclusions.

\section{Weak-field limit of $f(R,\lmatt)$ theory}
\label{sec:weak_field}

The $f(R,\lmatt)$ theory is defined by the action~\cite{HarkoLobo2010,BertolamiHarkoLoboParamos2008}
\begin{equation}
S=\int f(R,\lmatt)\sqrt{-g}\,\mathrm{d}^4x,
\label{eq:action}
\end{equation}
where $g$ denotes the determinant of the metric $g_{\mu\nu}$.

The null variation of the action \eqref{eq:action} with respect to the metric
leads to the following field equations~\cite{HarkoLobo2010}:
\begin{equation}
f_R R_{\mu\nu}
-\frac{1}{2}f g_{\mu\nu}
+\Delta f_Rg_{\mu\nu}
-f_{R;\mu\nu}
=
\frac{1}{2}f_{\lmatt}
\left(
T_{\mu\nu}-\lmatt g_{\mu\nu}
\right),
\label{eq:field_eq}
\end{equation}
where $\Delta := \nabla_\alpha\nabla^\alpha$ is the Laplace-Beltrami operator and we have defined
\begin{align}
f_R&:=\partial_R f,\\
f_{\lmatt}&:=\partial_{\lmatt} f,
\label{eq:f_derivatives}
\end{align}
The energy-momentum tensor of matter is defined by~\cite{LandauLifshitz}
\begin{equation}
T_{\mu\nu}
:=
-\frac{2}{\sqrt{-g}}
\frac{\delta\left(\sqrt{-g}\lmatt\right)}
{\delta g^{\mu\nu}}.
\label{eq:energy_momentum}
\end{equation}
Introducing the Einstein tensor,
\begin{equation}
G_{\mu\nu}
=
R_{\mu\nu}
-\frac{1}{2}g_{\mu\nu}R,
\label{eq:Einstein_tensor}
\end{equation}
equation \eqref{eq:field_eq} can be written equivalently as~\cite{HarkoLobo2010}
\begin{equation}
\begin{aligned}
G_{\mu\nu}
={}&
\frac{1}{f_R}
\Bigg[
\frac{1}{2}g_{\mu\nu}\left(f-Rf_R\right)
+f_{R;\mu\nu}
-g_{\mu\nu}\Delta f_R \\
&+\frac{1}{2}f_{\mathcal{L}_{\mathrm{matt}}}
\left(
T_{\mu\nu}-\mathcal{L}_{\mathrm{matt}}\, g_{\mu\nu}
\right)
\Bigg].
\label{eq:Einstein}
\end{aligned}
\end{equation}
Equations \eqref{eq:field_eq}--\eqref{eq:Einstein} constitute the starting point to study the weak-field regime of the theory. In what follows we develop the geometric and matter quantities according to their PN order.

\subsection{Post-Newtonian expansion}
\label{sec:PN_expansion}

To study the weak-field regime, we organize the geometric and matter
quantities by means of a PN expansion~\cite{LandauLifshitz,Chandrasekhar1965}. In what follows, the
superscript in parentheses indicates the perturbation order of 2 that particular quantity, in terms of powers of \( 1 / c \), where \( c \) is the speed of light. We adopt the traditional convention:
\begin{equation}
\mathcal{O}(n):=\mathcal{O}\left(c^{-n}\right).
\label{eq:order_definition}
\end{equation}
Thus, a quantity $X^{(n)}$ corresponds to a term of order
$\mathcal{O}(n)$. We use a metric  signature $(+,-,-,-)$ and work in
units such that $c=1$. 

Following~\cite{Wald1984}, for the case of a static and spherically symmetric configuration, we write
the line element as:
\begin{equation}
\text{d}s^2=g_{tt}(r)\,\text{d}t^2+g_{rr}(r)\,\text{d}r^2-r^2\text{d}\Omega^2,
\label{eq:metric_general}
\end{equation}
where $\text{d}\Omega^2=\text{d}\theta^2+\sin^2\theta\,\text{d}\phi^2$ represents the angular displacement.

In the weak-field regime, the metric is considered as a
perturbation about Minkowski space-time~\cite{LandauLifshitz,CapozzielloDeLaurentis2012}:
\begin{equation}
g_{\mu\nu}=\eta_{\mu\nu}+h_{\mu\nu},
\qquad
|h_{\mu\nu}|\ll1,
\label{eq:metric_perturbation}
\end{equation}
with
\begin{equation}
\eta_{\mu\nu}=\operatorname{diag}
\left(
1,-1,-r^2,-r^2\sin^2\theta
\right).
\label{eq:Minkowski_spherical}
\end{equation}
At the first non-trivial orders, the metric components are expanded as~\cite{CapozzielloDeLaurentis2012}
\begin{equation}
\left\{
\begin{aligned}
g_{tt}(r)
&=1+g_{tt}^{(2)}(r)+g_{tt}^{(4)}(r)
+\mathcal{O}(6),\\[0.5em]
g_{rr}(r)
&=-1+g_{rr}^{(2)}(r)
+\mathcal{O}(4),\\[0.5em]
g_{\theta\theta}(r)
&=-r^2,\\[0.5em]
g_{\phi\phi}(r)
&=-r^2\sin^2\theta.
\end{aligned}
\right.
\label{eq:metric_expansion}
\end{equation}
The perturbations $g_{tt}^{(2)}$ and $g_{rr}^{(2)}$ constitute the
first gravitational corrections to the Minkowski background.

Consistently with the expansion of the metric, the Ricci scalar
is expanded as
\begin{equation}
R=R^{(0)}+R^{(2)}+\mathcal{O}(4).
\label{eq:R_expansion}
\end{equation}
Since the background is a Minkowski space-time, it follows that:
\begin{equation}
R^{(0)}=0.
\label{R0}
\end{equation}
In what follows, we will only deal with perfect fluids~\cite{LandauLifshitz,Chandrasekhar1965} and so, the  matter Lagrangian has the following form~\cite{mendoza-silva_01, mendoza-silva_02, mendoza-silva_03}:
\begin{equation}
\lmatt=\rho+\xi,
\label{eq:Lm_fluid}
\end{equation}
where $\rho$ represents the energy density associated with the rest
mass and $\xi$ the pure thermodynamical internal energy density of the fluid. According
to the PN orders of approximation, the contribution associated with the rest mass
is of order $\mathcal{O}(-2)$, while the internal energy
contributes at order $\mathcal{O}(0)$~\cite{LandauLifshitz,Chandrasekhar1965}. In other words:
\begin{equation}
\lmatt=\lmatt^{(-2)}+\lmatt^{(0)},
\label{eq:Lm_expansion}
\end{equation}
with the identifications
\begin{align}
\lmatt^{(-2)}:=&\rho, \label{eq:Lm_rho}\\
\lmatt^{(0)}:=&\xi.
\label{eq:Lm_xi}
\end{align}
In what follows we will work directly with $\lmatt^{(-2)}$ and
$\lmatt^{(0)}$, instead of $\rho$ and $\xi$, in order to keep
explicit the PN order of each term during the perturbative
development. Once the solutions are obtained, we will recover the
physical notation in terms of the energy density $\rho$ and the
internal energy density $\xi$.

For the dominant contribution we adopt the decomposition
\begin{equation}
\lmatt^{(-2)}=\lmattstar^{(-2)}+\delta \lmatt^{(-2)},
\label{eq:Lm_minus2_perturbation}
\end{equation}
where $\lmattstar^{(-2)}:=\rho_*$ denotes the constant
reference background rest-mass energy density of the
matter sector, and $\delta\lmatt^{(-2)}$ represents
the spatial variation with respect to this reference
background. The physical background density will be
determined later from the solutions of the field equations.

Following the perturbative treatment employed in $f(R)$ 
gravity~\cite{CapozzielloStabileTroisi2009}, we expand $f(R,\lmatt)$ about
the reference values of $R$ and $\lmatt$. Introducing the perturbation of the matter Lagrangian as:
\begin{equation}
\begin{aligned}
\delta \lmatt&=\lmatt-\lmattstar^{(-2)} \\ &=\lmatt^{(-2)}+\lmatt^{(0)}-\lmattstar^{(-2)} \\ &=\delta \lmatt^{(-2)}+\lmatt^{(0)},
\label{eq:delta_Lm}
\end{aligned}
\end{equation}
the Taylor expansion of $f(R,\lmatt)$ can be organised, up to second perturbation order as
\begin{equation}
\label{taylor}
\begin{aligned}
f(R,\lmatt) =&f_0+f_1 R
+g_1 \delta \lmatt
+f_2 R^2
+h_{12} R\,\delta \lmatt  +\\ & g_2 (\delta \lmatt)^2
+\mathcal{O}(3),
\end{aligned}
\end{equation}
where we have defined
\begin{align}
f_0 &:= f(0, \lmattstar^{(-2)}), \\
f_1 &:= \left. \partial_R f \right|_{(0, \lmattstar^{(-2)})}, \\
g_1 &:= \left. \partial_{\lmatt} f \right|_{(0, \lmattstar^{(-2)})}, \\
f_2 &:= \frac{1}{2} \left. \partial^2_R f \right|_{(0, \lmattstar^{(-2)})}, \\
h_{12} &:= \left. \partial_R\partial_{\lmatt} f \right|_{(0, \lmattstar^{(-2)})}, \\
g_2 &:= \frac{1}{2} \left. \partial^2_{\lmatt} f \right|_{(0, \lmattstar^{(-2)})}.
\end{align}
Note that at $\mathcal{O}(2) $, the mixed term $h_{12}R\,\delta \lmatt$ describes 
the coupling between curvature and matter perturbations ~\cite{BertolamiHarkoLoboParamos2008}.

\subsection{Field equations by orders}
\label{sec:field_equations_orders}

From the expansion \eqref{taylor}, the different contributions
of $f(R,\lmatt)$ can be organised according to their PN order. This
decomposition allows an explicit identification of the terms
that enter the field equations at the required order, i.e.:
\begin{equation}
    f=f^{(-2)}+f^{(0)}+f^{(2)},
\end{equation}
where
\begin{align}
f^{(-2)}&=
g_1\delta \lmatt^{(-2)}
+2g_2\delta \lmatt^{(-2)}\lmatt^{(0)},
\label{eq:f_minus2}\\
f^{(0)}&=
f_0
+g_1\lmatt^{(0)}
+h_{12}R^{(2)}\delta \lmatt^{(-2)}
+g_2\left(\lmatt^{(0)}\right)^2,
\label{eq:f_0}\\
f^{(2)}&=
f_1R^{(2)}
+h_{12}R^{(2)}\lmatt^{(0)}.
\label{eq:f_2}
\end{align}
Similarly, the derivatives of $f$ with respect to $R$ are
decomposed as
\begin{equation}
     f_R=f^{(-2)}_R+f^{(0)}_R+f^{(2)}_R, \label{eq:f_R}
\end{equation}
where
\begin{align}
f_R^{(-2)}&=
h_{12}\delta \lmatt^{(-2)},
\label{eq:fR_minus2}\\
f_R^{(0)}&=
f_1+h_{12}\lmatt^{(0)},
\label{eq:fR_0}\\
f_R^{(2)}&=
2f_2R^{(2)},
\label{eq:fR_2}
\end{align}
while the derivatives with respect to $\lmatt$ are given by
\begin{equation}
     f_{\lmatt}=f^{(-2)}_{\lmatt}+f^{(0)}_{\lmatt}+f^{(2)}_{\lmatt},
\end{equation}
where
\begin{align}
f_{\lmatt}^{(-2)}&=
2g_2\delta \lmatt^{(-2)},
\label{eq:fLminus2}\\
f_{\lmatt}^{(0)}&=
g_1+2g_2\lmatt^{(0)},
\label{eq:fL_0}\\
f_{\lmatt}^{(2)}&=
h_{12}R^{(2)}.
\label{eq:fL_2}
\end{align}
For the case of a perfect fluid, the energy-momentum tensor is given by
~\cite{LandauLifshitz,Chandrasekhar1965,mendoza-silva_01,mendoza-silva_02,mendoza-silva_03}: 
\begin{equation}
T_{\mu\nu}
=
(\lmatt+p)u_\mu u_\nu
-pg_{\mu\nu},
\label{eq:T_mu_nu}
\end{equation}
where $u^\mu$ is the four-velocity of the fluid and $p$ its pressure, which contributes at order $\mathcal{O}(0)$.

For a fluid at rest in the coordinate system that we are considering, the
four-velocity is expanded, at the required order, as
\begin{equation}
u_\mu
=
u_\mu^{(0)}+u_\mu^{(2)}+\mathcal{O}(4),
\label{eq:u_expansion}
\end{equation}
with
\begin{align}
&u_\mu^{(0)}=
(1,0,0,0), \label{eq:u_orders_cero}\\
&u_\mu^{(2)}=
\left(
\frac{1}{2}g_{tt}^{(2)},0,0,0
\right),
\label{eq:u_orders}
\end{align}
according to the normalisation condition
$u_\mu u^\mu=1$ ~\cite{Wald1984}.

Consistently, the energy-momentum tensor is organised according to
its PN order as ~\cite{LandauLifshitz,Chandrasekhar1965}
\begin{equation}
T_{\mu\nu}
=
T_{\mu\nu}^{(-2)}
+
T_{\mu\nu}^{(0)}
+
T_{\mu\nu}^{(2)},
\label{eq:T_expansion}
\end{equation}
where
\begin{align}
T_{\mu\nu}^{(-2)}=&
\lmatt^{(-2)}\,u_\mu^{(0)}u_\nu^{(0)},
\label{eq:T_minus2}\\
T_{\mu\nu}^{(0)}=&
\lmatt^{(-2)}
\left(
u_\mu^{(0)}u_\nu^{(2)}
+
u_\mu^{(2)}u_\nu^{(0)}
\right) +\nonumber \\ 
&
(\lmatt^{(0)}+p)u_\mu^{(0)}u_\nu^{(0)}
-pg_{\mu\nu}^{(0)},
\label{eq:T_0}\\
T_{\mu\nu}^{(2)}=&
(\lmatt^{(0)}+p)
\left(
u_\mu^{(0)}u_\nu^{(2)}
+
u_\mu^{(2)}u_\nu^{(0)}
\right)
-pg_{\mu\nu}^{(2)}.
\label{eq:T_2}
\end{align}
The dominant contribution, $T_{\mu\nu}^{(-2)}$, is determined by
the energy density associated with the rest mass, while the
following orders incorporate the internal energy, the pressure, and the
corrections induced by the gravitational perturbation
~\cite{LandauLifshitz,Chandrasekhar1965}.

The trace of the energy-momentum tensor, which is needed in the field
equations, is decomposed as
\begin{equation}
T=g^{\mu\nu}T_{\mu\nu} =
T^{(-2)}+T^{(0)},
\label{eq:T_trace}
\end{equation}
with
\begin{align}
T^{(-2)}&=\rho,\label{T-2}\\
T^{(0)}&=\xi-3p.
\label{T0}
\end{align}
This specifies the matter content necessary for the development that follows.

\subsection{Linearised field equations}

To linearize the field equations \eqref{eq:Einstein}, we expand $f_R$ and its inverse. Considering (\ref{eq:f_R}) where the different contributions are given by \eqref{eq:fR_minus2}--\eqref{eq:fR_2}, and assuming that $f_R^{(-2)}$ is the dominant contribution, we obtain, at first order in the perturbative ratios,
\begin{equation}
\frac{1}{f_R}
\simeq
\frac{1}{f_R^{(-2)}}
\left(1-\frac{f_R^{(0)}}{f_R^{(-2)}}
-\frac{f_R^{(2)}}{f_R^{(-2)}}\right),
\label{eq:inv_fR}
\end{equation}
under the conditions $f_R^{(-2)}\neq0$ and
\begin{equation}
\left|\frac{f_R^{(0)}}{f_R^{(-2)}}\right|\ll1,
\qquad
\left|\frac{f_R^{(2)}}{f_R^{(-2)}}\right|\ll1.
\label{eq:fR_conditions}
\end{equation}
Substituting this expression, together with the expansions of $f$, $f_{\lmatt}$ and $T_{\mu\nu}$, into
\eqref{eq:Einstein}, and retaining the contributions compatible with the PN hierarchy and the perturbative approximation adopted, we obtain the field equations corresponding to each order.

At order $\mathcal{O}(-2)$,
\begin{equation}
G_{\mu\nu}^{(-2)}
=
\frac{f_{\lmatt}^{(-2)}}{2f_R^{(-2)}}
\left(
T_{\mu\nu}^{(-2)}
-
\lmatt^{(-2)}g_{\mu\nu}^{(0)}
\right).
\label{eq:G_minus2}
\end{equation}
The geometric expansion does not contain a contribution of this order,
so $G_{\mu\nu}^{(-2)}=0$. Since
$f_R^{(-2)}\neq0$, one obtains
\begin{equation}
f_{\lmatt}^{(-2)}
\left(
T_{\mu\nu}^{(-2)}
-
\lmatt^{(-2)}g_{\mu\nu}^{(0)}
\right)
=
0.
\label{eq:consistency_minus2}
\end{equation}
Using \eqref{eq:T_minus2} we have,
\begin{equation}
T_{\mu\nu}^{(-2)}
-
\lmatt^{(-2)}g_{\mu\nu}^{(0)}
=
\lmatt^{(-2)}
\left(
u_\mu^{(0)}u_\nu^{(0)}
-
g_{\mu\nu}^{(0)}
\right).
\label{eq:matter_minus2}
\end{equation}
For the fluid at rest considered, this expression does not vanish
identically and, therefore,
\begin{equation}
f_{\lmatt}^{(-2)}=0.
\label{fLm-2=0}
\end{equation}
For a non-trivial matter configuration, $\delta \lmatt^{(-2)}\neq0$, equation \eqref{eq:fLminus2} implies
\begin{equation}
g_2=0.
\label{eq:g2_zero}
\end{equation}
Consequently, consistency at order $\mathcal{O}(-2)$ eliminates
the quadratic term in $\lmatt$ from the expansion considered.

At order $\mathcal{O}(0)$, condition \eqref{eq:g2_zero} implies equation~\eqref{fLm-2=0}
and so, the field equations \eqref{eq:Einstein} reduce to
\begin{equation}
\begin{aligned}
G_{\mu\nu}^{(0)}
={}&
\frac{1}{f_R^{(-2)}}
\left[
\frac{1}{2}g_{\mu\nu}^{(0)}f^{(-2)}
+f_{R,\nu\mu}^{(-2)}
-\Gamma^{r\,(0)}_{\mu\nu}f_{R,r}^{(-2)}-
\right.\\
&\left.
g_{\mu\nu}^{(0)}\Delta f_R^{(-2)}
+\frac{1}{2}f_{\lmatt}^{(0)}
\left(
T_{\mu\nu}^{(-2)}
-\lmatt^{(-2)}g_{\mu\nu}^{(0)}
\right)
\right].
\end{aligned}
\label{eq:G0}
\end{equation}
Since the geometric background is flat, $G_{\mu\nu}^{(0)}=0$. For the static configuration considered, the time derivatives
vanish and, at the required order, the Laplace-Beltrami operator reduces to $\Delta\simeq-\nabla^2$. Taking then the trace of \eqref{eq:G0} and using \eqref{eq:delta_Lm}, \eqref{eq:fR_minus2}, \eqref{eq:fL_0} and \eqref{eq:T_minus2}, one obtains
\begin{equation}
\nabla^2\lmatt^{(-2)}
-m^2\lmatt^{(-2)}=-4m^2\lmattstar^{(-2)},
\label{eq:klein_gordon}
\end{equation}
where
\begin{equation}
m^2:=-\frac{g_1}{6h_{12}}>0.
\label{eq:m2}
\end{equation}
Therefore, the condition $m^2>0$ requires  that $g_1$ and $h_{12}$
have opposite signs.

For a spherically symmetric configuration, the solution of
the Yukawa-like equation~\eqref{eq:klein_gordon} is
\begin{equation}
\lmatt^{(-2)}=4\lmattstar^{(-2)}+A\frac{e^{-mr}}{r},
\label{eq:Lm_solution}
\end{equation}
where $A$ is an integration constant. Recalling (\ref{eq:Lm_rho}), we define
\begin{equation}
\rho_0:= 4\lmattstar^{(-2)},
\label{eq:rho0_definition}
\end{equation}
so that the energy density associated with the rest mass
adopts the form
\begin{equation}
\rho(r)
=
\rho_0
+
A\frac{e^{-mr}}{r}.
\label{eq:rho_solution}
\end{equation}
Thus, $\rho_0$ represents the background value of the density, while
the second term introduces a radial Yukawa-type contribution.

At order $\mathcal{O}(2)$, using \eqref{eq:g2_zero} and the
condition $G_{\mu\nu}^{(0)}=0$, the linearised Einstein tensor
adopts the form
\begin{equation}
G_{\mu\nu}^{(2)}
=
R_{\mu\nu}^{(2)}
-\frac{1}{2}g_{\mu\nu}^{(0)}R^{(2)}.
\label{eq:G2_def}
\end{equation}
The field equation \eqref{eq:Einstein} then becomes
\begin{equation}
\begin{aligned}
R_{\mu\nu}^{(2)}
={}&
\frac{1}{f_R^{(-2)}}\Biggl[
g_{\mu\nu}^{(0)}
\left(
\frac{f^{(0)}}{2}
-\Delta f_R^{(0)}
\right)+\\
&
g_{\mu\nu}^{(2)}
\left(
\frac{f^{(-2)}}{2}
-\Delta f_R^{(-2)}
\right)+
\\
&
f_{R,\nu\mu}^{(0)}
-\Gamma^{r\,(0)}_{\mu\nu}f_{R,r}^{(0)}
-\Gamma^{r\,(2)}_{\mu\nu}f_{R,r}^{(-2)}+\\
&
\frac{f_{\lmatt}^{(2)}}{2}
\left(
T_{\mu\nu}^{(-2)}
-\lmatt^{(-2)}g_{\mu\nu}^{(0)}
\right)+
\\
&
\frac{f_{\lmatt}^{(0)}}{2}
\left(
T_{\mu\nu}^{(0)}
-\lmatt^{(-2)}g_{\mu\nu}^{(2)}
-\lmatt^{(0)}g_{\mu\nu}^{(0)}
\right)
\Biggr].
\end{aligned}
\label{eq:G2}
\end{equation}
For the static and spherically symmetric metric considered, the
non-zero components of the Ricci tensor at the required order are
\begin{align}
R_{tt}^{(2)}
&=
\frac{1}{2}g_{tt,rr}^{(2)}
+\frac{1}{r}g_{tt,r}^{(2)},
\label{eq:Rtt}\\
R_{rr}^{(2)}
&=
-\frac{1}{2}g_{tt,rr}^{(2)}
-\frac{1}{r}g_{rr,r}^{(2)},
\label{eq:Rrr}\\
R_{\theta\theta}^{(2)}
&=
-g_{rr}^{(2)}
-\frac{r}{2}g_{rr,r}^{(2)}
-\frac{r}{2}g_{tt,r}^{(2)},
\label{eq:Rthetatheta}\\
R_{\phi\phi}^{(2)}
&=
\sin^2\theta\,R_{\theta\theta}^{(2)}.
\label{eq:Rphiphi}
\end{align}
Substituting these expressions into \eqref{eq:G2}, together with the  previously
obtained expansions, the independent components
of the field equations take the following form:

For the $tt$ component,
\begin{equation}
\begin{aligned}
\nabla^2g_{tt}^{(2)}
=&
\frac{f_0+g_1\lmatt^{(0)}}
{h_{12}\delta \lmatt^{(-2)}}
+R^{(2)}
+\frac{2\nabla^2\lmatt^{(0)}}
{\delta \lmatt^{(-2)}}+ \\
&
\left(
\frac{g_1}{h_{12}}
+
\frac{2\nabla^2\lmatt^{(-2)}}
{\delta \lmatt^{(-2)}}
\right)g_{tt}^{(2)}
-
\frac{\mathcal{L}_{\text{matt}_{,r}}^{(-2)}}
{\delta \lmatt^{(-2)}}
g_{tt,r}^{(2)}.
\end{aligned}
\label{eq:tt_exact}
\end{equation}
The $rr$ component leads to
\begin{equation}
\begin{aligned}
&\frac{1}{2}g_{tt,rr}^{(2)}
+
\left(
\frac{1}{r}
+
\frac{\mathcal{L}_{\text{matt}_{,r}}^{(-2)}}
{2\delta \lmatt^{(-2)}}
\right)
g_{rr,r}^{(2)}
=\\
&\frac{f_0-g_1p}
{2h_{12}\delta \lmatt^{(-2)}}+
\frac{1}{2}
\left(
1-\frac{\lmatt^{(-2)}}
{\delta \lmatt^{(-2)}}
\right)R^{(2)}
+
\frac{2\mathcal{L}_{\text{matt}_{,rr}}^{(0)}}
{\delta \lmatt^{(-2)}}-
\\
&
\left(
\frac{g_1}{2h_{12}}
-\frac{g_1\lmatt^{(-2)}}
{2h_{12}\delta \lmatt^{(-2)}}
+\frac{\nabla^2\lmatt^{(-2)}}
{\delta \lmatt^{(-2)}}
\right)g_{rr}^{(2)}.
\end{aligned}
\label{eq:rr_exact}
\end{equation}
For its part, the $\theta\theta$ component is
\begin{equation}
\begin{aligned}
&\left(
\frac{1}{r}
+
\frac{\mathcal{L}_{\text{matt}_{,r}}^{(-2)}}
{\delta \lmatt^{(-2)}}
\right)
g_{rr}^{(2)}
+\frac{1}{2}g_{rr,r}^{(2)}
+\frac{1}{2}g_{tt,r}^{(2)}
=\\
&\frac{r(f_0-g_1p)}
{2h_{12}\delta \lmatt^{(-2)}}
+
\frac{r}{2}
\left(
1-\frac{\lmatt^{(-2)}}
{\delta \lmatt^{(-2)}}
\right)R^{(2)}+\\
&
\frac{1}{\delta \lmatt^{(-2)}}
\left(
r\mathcal{L}_{\text{matt}_{,rr}}^{(0)}
+\mathcal{L}_{\text{matt}_{,r}}^{(0)}
\right).
\end{aligned}
\label{eq:theta_exact}
\end{equation}
The trace at order $\mathcal{O}(0)$ provides the equation for
$\lmatt^{(0)}$:
\begin{equation}
\begin{aligned}
\nabla^2\lmatt^{(0)}
-m^2\lmatt^{(0)}
=&
-\frac{1}{3}
\left(
\delta \lmatt^{(-2)}
-\frac{3}{2}\lmatt^{(-2)}
\right)R^{(2)} +\\
&
\frac{g_1p}{2h_{12}}
-\frac{2f_0}{3h_{12}},
\end{aligned}
\label{eq:L0_exact}
\end{equation}
while the trace at order $\mathcal{O}(2)$ determines the equation
for the curvature,
\begin{equation}
6f_2\nabla^2R^{(2)}
+
\left(
f_1
-\frac{1}{2}h_{12}\lmatt^{(0)}
-\frac{3}{2}h_{12}p(r)
\right)R^{(2)}
=
0.
\label{eq:R_exact}
\end{equation}
In the following Section we simplify this system of equations under the approximations corresponding to the appropriate regime.

\subsubsection{Separation between background and perturbations}

The solution obtained for $\lmatt^{(-2)}$ is given by
\eqref{eq:Lm_solution}. Consequently, its deviation from the
reference value is
\begin{equation}
\delta \lmatt^{(-2)}
=
3\lmattstar^{(-2)}
+
A\frac{e^{-mr}}{r}.
\label{eq:deltaLminus2}
\end{equation}
To quantify the amplitude of the radial contribution with respect to the constant term, we define
\begin{equation}
\epsilon_{-2}
:=
\left|
\frac{Ae^{-mr}}
{3\lmattstar^{(-2)}r}
\right|.
\label{eq:epsilon_minus2}
\end{equation}
In the regime $\epsilon_{-2}\ll1$, the radial contribution
constitutes a small perturbation with respect to the dominant term of
$\delta \lmatt^{(-2)}$. Indeed, \eqref{eq:deltaLminus2} can be written as
\begin{equation}
\delta \lmatt^{(-2)}
=
3\lmattstar^{(-2)}
\left(
1+
\frac{Ae^{-mr}}
{3\lmattstar^{(-2)}r}
\right),
\end{equation}
so that
\begin{equation}
\delta \lmatt^{(-2)}
=
3\lmattstar^{(-2)}
\left(
1+\mathcal O(\epsilon_{-2})
\right).
\end{equation}
Its inverse satisfies, at the same order,
\begin{equation}
\frac{1}{\delta \lmatt^{(-2)}}=\frac{1}{3\lmattstar^{(-2)}}
\left(
1-\mathcal O(\epsilon_{-2})
\right).
\end{equation}
Therefore, at dominant order,
\begin{align}
&\lmatt^{(-2)}
\simeq
4\lmattstar^{(-2)},\\
&\delta \lmatt^{(-2)}
\simeq
3\lmattstar^{(-2)}.
\label{eq:Lminus2_approx}
\end{align}
The derivatives that appear in equations~\eqref{eq:tt_exact}-\eqref{eq:theta_exact} may be neglected when, in addition to
$\epsilon_{-2}\ll1$, they are subdominant as compared to the other terms 
in the field equations.

In particular, this approximation does not require imposing $mr\gg1$; the
relevant condition on the amplitude of the matter perturbation is
$\epsilon_{-2}(r)\ll1$.

\subsubsection{Matter sector perturbations}

For the zero-order matter sector, an analogous
separation is introduced,
\begin{align}
\lmatt^{(0)}
&=\mathcal{L}_{\mathrm{matt}0}^{(0)}
+\delta\lmatt^{(0)},
\label{eq:L_background}\\
p&=p_0+\delta p,
\label{eq:p_background}
\end{align}
where $\mathcal{L}_{\mathrm{matt}0}^{(0)}$ and
$p_0$ represent the constant background values
of the internal energy density and pressure,
respectively, while $\delta\lmatt^{(0)}$
and $\delta p$ denote their radial corrections.

We assume that the radial corrections vanish
at large distances,
\begin{align}
\lim_{r\to\infty}\delta\lmatt^{(0)}(r)&=0,\label{eq:matter_smallness_xi}\\
\lim_{r\to\infty}\delta p(r)&=0\label{eq:matter_smallness_p}.
\end{align}
Consequently, at sufficiently large radial
distances, the matter variables approach
their constant background values,
\begin{align}
\lmatt^{(0)}(r)
&\simeq\mathcal{L}_{\mathrm{matt}0}^{(0)},
\label{eq:matter_asymptotic_approximation_xi}\\
p(r)&\simeq p_0.
\label{eq:matter_asymptotic_approximation_p}
\end{align}
These asymptotic conditions do not imply that
the radial corrections must be eliminated
from the field equations. On the contrary,
they are retained when they constitute
the quantities to be determined.

\subsubsection{Gravitational scale and quasi-Newtonian regime}
%============================================================

In addition to the weak-field condition, $|g_{\mu\nu}^{(2)}|\ll1$, we consider a quasi-Newtonian (QN) regime, in which the dominant differential structure of the Newtonian limit is preserved, but corrections to the gravitational potential are admitted, such as the Yukawa-type ones obtained in modified gravity theories ~\cite{Stelle1978,CapozzielloStabileTroisi2009}.

In our case, this regime is established by considering that the algebraic contributions proportional to $m^2$ are small compared to the characteristic differential scales of the metric perturbations. 

At the dominant order in $\epsilon_{-2}$,
using \eqref{eq:Lminus2_approx},
the metric part of the $tt$ component
adopts the form
\begin{equation}
\begin{aligned}
&\nabla^2g_{tt}^{(2)}
-
\left(
\frac{g_1}{h_{12}}
+
\frac{2\nabla^2\lmatt^{(-2)}}
{\delta \lmatt^{(-2)}}
\right)g_{tt}^{(2)}
+
\frac{\mathcal{L}_{\text{matt}_{,r}}^{(-2)}}
{\delta \lmatt^{(-2)}}
g_{tt,r}^{(2)}\\
&\simeq
\nabla^2g_{tt}^{(2)}
+
6m^2g_{tt}^{(2)}.
\end{aligned}
\label{eq:tt_quasi_newtonian_reduction}
\end{equation}
Similarly, the $rr$ component
reduces to
\begin{equation}
\begin{aligned}
&g_{tt,rr}^{(2)}+
\left(
\frac{g_1}{h_{12}}
-\frac{g_1\lmatt^{(-2)}}
{h_{12}\delta \lmatt^{(-2)}}
+
\frac{2\nabla^2\lmatt^{(-2)}}
{\delta \lmatt^{(-2)}}
\right)g_{rr}^{(2)}+\\
&
\left(
\frac{2}{r} +
\frac{\mathcal{L}_{\text{matt}_{,r}}^{(-2)}}
{\delta \lmatt^{(-2)}}
\right)g_{rr,r}^{(2)}
\simeq
g_{tt,rr}^{(2)}
+\frac{2}{r}g_{rr,r}^{(2)}
+2m^2g_{rr}^{(2)}.
\end{aligned}
\label{eq:rr_quasi_newtonian_reduction}
\end{equation}
% ============================================================
% Proposed replacement: QN regime, system, and solutions
% ============================================================
Let $\mathcal I=[r_{\min},r_{\max}]\subset(0,\infty)$ be a compact radial interval.
We introduce a characteristic length $\ell_\text{g}>0$ and the dimensionless
variable
\begin{equation}
x:=\frac{r}{\ell_\text{g}},
\text{ defined on } \mathcal J:=\left[\frac{r_{\min}}{\ell_\text{g}},
\frac{r_{\max}}{\ell_\text{g}}\right].
\label{eq:qn_dimensionless_radius}
\end{equation}
Here $\mathcal J$ is the image of $\mathcal I$ under the radial change of
variable. We define
\begin{equation}
\epsilon_\text{g}:=6m^2\ell_\text{g}^2
=\left|\frac{g_1}{h_{12}}\right|\ell_\text{g}^2,
\label{eq:epsilon_g}
\end{equation}
The QN hierarchy is
\begin{equation}
\epsilon_\text{g}\ll1,
\qquad
\ell_\text{g}\ll\frac{1}{\sqrt{6}\,m}.
\label{eq:quasi_newtonian_hierarchy}
\end{equation}
The two inequalities are equivalent. The length $\ell_\text{g}$ must
be related to the characteristic differential scale of the problem;
the approximation cannot be justified by choosing it arbitrarily
small without considering the interval, the boundary conditions, and
the scales over which the solutions vary.

In general terms, the field equations can possess undamped oscillatory solutions, which are unphysical. To 
find physical solutions to the problem, we proceed as follows.

For a continuous function $F$, we define the supremum norm
\begin{equation}
\|F\|_{\infty,\mathcal I}:=
\sup_{r\in\mathcal I}|F(r)|.
\label{eq:qn_sup_norm}
\end{equation}
We denote the metric solutions of the reduced system by
$\bar g_{tt}^{(2)}$ and $\bar g_{rr}^{(2)}$ and introduce
\begin{equation}
\begin{aligned}
G_t&:=\|\bar g_{tt}^{(2)}\|_{\infty,\mathcal I},\\
G_r&:=\|\bar g_{rr}^{(2)}\|_{\infty,\mathcal I}.
\end{aligned}
\label{eq:qn_metric_amplitudes}
\end{equation}
where $\bar g_{tt}^{(2)}$ and $\bar g_{rr}^{(2)}$
denote the temporal and radial metric solutions
of the reduced QN system, respectively. 

These amplitudes are finite by continuity on the compact interval
$\mathcal I$. For $G_t>0$, we write
\begin{equation}
\bar g_{tt}^{(2)}(r)=G_tu(x),
\quad
|u(x)|\leq1,
\label{eq:qn_metric_profiles_t}
\end{equation}
and for $G_r>0$:
\begin{equation}
  \bar   g_{rr}^{(2)}(r)=G_rv(x), \quad |v(x)|\leq1.
\label{eq:qn_metric_profiles_r}
\end{equation}
The functions $u$ and $v$ are dimensionless profiles. In particular,
$G_t\ll1$ and $G_r\ll1$ express the weak-field condition, but
do not imply that $|u|\ll1$ or $|v|\ll1$.

To transform the remaining equations, we introduce
\begin{equation}
\begin{aligned}
R^{(2)}(r)&=\frac{G_t}{\ell_\text{g}^2}\mathcal R(x),\\
\lmatt^{(0)}(r)&=L_0\eta(x),\\
p(r)&=P_0q(x),
\end{aligned}
\label{eq:qn_matter_profiles}
\end{equation}
where $\mathcal R$ is a dimensionless curvature and
$L_0:=\|\lmatt^{(0)}\|_{\infty,\mathcal I}$,
$P_0:=\|p\|_{\infty,\mathcal I}$ are positive scales of the
internal energy and pressure, respectively and since
$L_0,P_0>0$, we then  have $|\eta|\leq1$ and $|q|\leq1$.

The chain rule yields
\begin{align}
\nabla^2 \bar g_{tt}^{(2)}
&=\frac{G_t}{\ell_\text{g}^2}\nabla_x^2u, \label{eq:qn_derivative_transformations_g}\\
\nabla^2\lmatt^{(0)}
&=\frac{L_0}{\ell_\text{g}^2}\nabla_x^2\eta, \label{eq:qn_derivative_transformations_l}
\end{align}
where $\nabla_x^2$ is the dimensionless radial Laplacian.

Before removing the algebraic terms, the approximate metric
equations under consideration are
\begin{equation}
\left\{
\begin{aligned}
&\nabla^2\bar g_{tt}^{(2)}+6m^2\bar g_{tt}^{(2)}-R^{(2)}
=\frac{f_0+g_1\lmatt^{(0)}}
{3h_{12}\lmattstar^{(-2)}}
+\frac{2\nabla^2\lmatt^{(0)}}{3\lmattstar^{(-2)}},\\[0.6em]
&
\bar g_{tt,rr}^{(2)}+\frac{2}{r}\bar g_{rr,r}^{(2)}
+2m^2\bar g_{rr}^{(2)}+\frac13R^{(2)} = 
\\
& \qquad \qquad  \frac{f_0-g_1p}{3h_{12}\lmattstar^{(-2)}}
+\frac{4\mathcal{L}_{\text{matt}_{,r}}^{(0)}}{3r\lmattstar^{(-2)}}.
\end{aligned}
\right.
\label{eq:qn_metric_equations_before_truncation}
\end{equation}
These equations already incorporate the preceding approximations concerning
the dominant matter contribution; therefore, the residuals calculated
below quantify exclusively the omission of the algebraic metric terms,
and not the total error relative to the field equations prior to
those approximations.

Substituting the previous transformations, we obtain the dimensionless temporal equation
\begin{equation}
\begin{aligned}
\nabla_x^2u+\epsilon_\text{g}u
-\mathcal R=&
\frac{f_0\ell_\text{g}^2}
{3h_{12}\lmattstar^{(-2)}G_t}
-\frac{\epsilon_\text{g}L_0}
{3\lmattstar^{(-2)}G_t}\eta +\\ &\frac{2L_0}{3\lmattstar^{(-2)}G_t}\nabla_x^2\eta.
\end{aligned}
\label{eq:qn_tt_dimensionless}
\end{equation}
Since $|u|\leq1$, for every $x\in\mathcal J$ we have
\begin{equation}
0\leq|\epsilon_\text{g} u|\leq\epsilon_\text{g},
\text{ and so }
\|\epsilon_\text{g} u\|_{\infty,\mathcal J}\leq\epsilon_\text{g}.
\label{eq:qn_tt_uniform_bound}
\end{equation}
In a family of configurations with $\epsilon_\text{g}\to0$, the
normalisation of $u$ is valid for each member of the family,
even if the profile depends on $\epsilon_\text{g}$. By the squeeze
theorem,
\begin{equation}
\lim_{\epsilon_\text{g}\to0}
\|\epsilon_\text{g}u\|_{\infty,\mathcal J}=0.
\label{eq:qn_tt_uniform_limit}
\end{equation}
Thus, the temporal algebraic metric term converges uniformly
to zero in the normalised equation.

For the radial component, the exact transformation gives:
\begin{equation}
\begin{aligned}
&u_{,xx}+\frac{2G_r}{G_tx}v_{,x}
+\frac{\epsilon_\text{g}G_r}{3G_t}v
+\frac13\mathcal R= \\ &
\frac{f_0\ell_\text{g}^2}
{3h_{12}\lmattstar^{(-2)}G_t}+\frac{\epsilon_\text{g}P_0}
{3\lmattstar^{(-2)}G_t}q
+\frac{4L_0}{3\lmattstar^{(-2)}G_t}\frac{\eta_{,x}}{x}.
\end{aligned}
\label{eq:qn_rr_dimensionless}
\end{equation}
The normalisation $|v|\leq1$ implies
\begin{equation}
\left\|\frac{\epsilon_\text{g}G_r}{3G_t}v
\right\|_{\infty,\mathcal J}
\leq\frac{\epsilon_\text{g}G_r}{3G_t}.
\label{eq:qn_rr_uniform_bound}
\end{equation}
Therefore, when using $G_t/\ell_\text{g}^2$ as the reference scale
for the radial equation, it is sufficient to require
$\epsilon_\text{g}G_r/G_t\to0$. In particular, this condition holds
if $G_r/G_t$ remains uniformly bounded in the considered solutions.

To avoid imposing an unnecessary relation between the amplitudes,
we introduce the common metric scale
\begin{equation}
G_{\mathrm{met}}:=\max\{G_t,G_r\}>0.
\label{eq:qn_common_amplitude}
\end{equation}
With this choice, both perturbations are less than or equal to
$G_{\mathrm{met}}$ in absolute value. Let $\mathscr R_{tt}$ and $\mathscr R_{rr}$
be the residuals obtained by substituting the reduced solutions into
the metric equations before omitting the algebraic terms:
\begin{equation}
\mathscr R_{tt}:=6m^2\bar g_{tt}^{(2)},\quad
\mathscr R_{rr}:=2m^2\bar g_{rr}^{(2)}.
\label{eq:metric_residuals}
\end{equation}
Their magnitudes satisfy
\begin{equation}
\begin{aligned}
\|\mathscr R_{tt}\|_{\infty,\mathcal I}&\leq6m^2G_t,\\
\|\mathscr R_{rr}\|_{\infty,\mathcal I}&\leq2m^2G_r.
\end{aligned}
\label{eq:qn_dimensional_residual_bounds}
\end{equation}
Normalising both equations by the same differential scale
$G_{\mathrm{met}}/\ell_\text{g}^2$, we obtain:
\begin{align}
\frac{\|\mathscr R_{tt}\|_{\infty,\mathcal I}}
{G_{\mathrm{met}}/\ell_\text{g}^2}&\leq\epsilon_\text{g},
\label{eq:metric_residual_t}\\
\frac{\|\mathscr R_{rr}\|_{\infty,\mathcal I}}
{G_{\mathrm{met}}/\ell_\text{g}^2}&\leq\frac{\epsilon_\text{g}}{3}.
\label{eq:metric_residual_r}
\end{align}
Consequently, both normalised metric residuals converge
uniformly to zero as $\epsilon_\text{g}\to0$. The bounds remain
valid even at points where individual derivatives vanish;
they do not require division by differential terms that might
cancel out. The common scale removes the need to control
$G_r/G_t$ for this residual estimate, although it does not replace
the stability conditions of the coupled system.

It is important to mention that removing $\epsilon_\text{g} u$ does not imply
removing all terms proportional to $g_1$. In
particular, the normalised matter terms contain the
combinations $
{\epsilon_\text{g}L_0}/(
{3|\lmattstar^{(-2)}|G_t})$,
$
{\epsilon_\text{g}P_0}/(
{3|\lmattstar^{(-2)}|G_t}),
%\label{eq:qn_matter_amplitude_ratios}
$
which do not necessarily tend to zero if the ratios $L_0/G_t$ or
$P_0/G_t$ vary with the QN parameters. Likewise, $|\eta|\leq1$ does not
by itself control $\eta_{,x}$ or $\nabla_x^2\eta$. Therefore, the
matter terms are retained in the reduced equations, including
$m^2\lmatt^{(0)}$ in the matter equation. This choice
formally preserves their characteristic scale, but the
compatibility of the matter and geometric balances must
be verified for each configuration.

The preceding bounds demonstrate the smallness of the normalised
residuals of the metric equations. They do not, by themselves,
prove that solutions of the full system converge to those of the
reduced system: such a conclusion would require
compatible boundary conditions and stability estimates
for the coupled problem. The QN approximation is understood here as
a locally controlled reduction in the residual sense, within the
domain where the adopted hierarchies are satisfied.

From this point onward, for notational simplicity, we omit the bar and 
denote by \(g_{tt}^{(2)}\) and \(g_{rr}^{(2)}\) the solutions of the reduced system.

\section{Analytical solutions}
\label{sec:analytical_solutions}

The preceding hierarchy leads to the reduced system
\begin{equation}
\left\{
\begin{aligned}
&\nabla^2g_{tt}^{(2)}-R^{(2)}
=\frac{f_0+g_1\lmatt^{(0)}}
{3h_{12}\lmattstar^{(-2)}}
+\frac{2\nabla^2\lmatt^{(0)}}{3\lmattstar^{(-2)}},\\[0.8em]
&g_{tt,rr}^{(2)}+\frac2r g_{rr,r}^{(2)}+\frac13R^{(2)}
=\frac{f_0-g_1p}{3h_{12}\lmattstar^{(-2)}}
+\frac{4\mathcal{L}_{\text{matt}_{,r}}^{(0)}}{3r\lmattstar^{(-2)}},\\[0.8em]
&\frac1r g_{tt,r}^{(2)}+\frac1r g_{rr,r}^{(2)}
+\frac2{r^2}g_{rr}^{(2)}+\frac13R^{(2)}
=\frac{f_0-g_1p}{3h_{12}\lmattstar^{(-2)}}+\\
&\frac2{3\lmattstar^{(-2)}}
\left(\mathcal{L}^{(0)}_{\text{matt}_{,rr}}+\frac{1}{r} \mathcal{L}^{(0)}_{\text{matt}_{,r}}\right),\\[0.8em]
&2g_{rr}^{(2)}+r\left(2g_{tt,r}^{(2)}-rR^{(2)}
+2g_{rr,r}^{(2)}+rg_{tt,rr}^{(2)}\right)=0,\\[0.8em]
&\nabla^2\lmatt^{(0)}-m^2\lmatt^{(0)}
=\lmattstar^{(-2)}R^{(2)}
+\frac{g_1p}{2h_{12}}-\frac{2f_0}{3h_{12}},\\[0.8em]
&\nabla^2R^{(2)}-\mu^2R^{(2)}=0.
\end{aligned}
\right.
\label{eq:system_final}
\end{equation}
Here $\mu$ is the inverse scale associated with the curvature mode,
with
\begin{equation}
\mu^2=-\frac1{6f_2}\left(f_1-
\frac12h_{12}\mathcal{L}_{\text{matt}0}^{(0)}-
\frac32h_{12}p_0\right)>0.
\label{eq:mu2_general}
\end{equation}
\subsection{General solutions}
\label{sec:general_solutions}
For $r>0$, $m>0$, $\mu>0$, and $\mu^{2}\neq m^{2}$, the exact
analytical solutions of the reduced system are
\begin{equation}
\left\{
\begin{aligned}
&\begin{aligned}
&g_{tt}^{(2)}(r)=\delta_0-\frac{C_\text{N}}{r}
+\frac{1}{6\lmattstar^{(-2)}}
\left(\frac{f_0}{3h_{12}}+2K\right)r^2-\\ &\frac{4\delta_2}{3\lmattstar^{(-2)}}\frac{e^{-mr}}r
+\frac{\delta_1(\mu^2+3m^2)}
{3\mu^2(\mu^2-m^2)}\frac{e^{-\mu r}}r,
\end{aligned}\\[0.6em]
&\begin{aligned}
&g_{rr}^{(2)}(r)=-\frac{C_\text{N}}{r}
-\frac{1}{6\lmattstar^{(-2)}}
\left(\frac{f_0}{3h_{12}}+2K\right)r^2-\\ &\frac{2\delta_2}{3\lmattstar^{(-2)}}
(1+mr)\frac{e^{-mr}}r+\\
&\frac{\delta_1(3m^2-\mu^2)}
{3\mu^2(\mu^2-m^2)}(1+\mu r)\frac{e^{-\mu r}}r,
\end{aligned}\\[0.6em]
&R^{(2)}(r)=\delta_1\frac{e^{-\mu r}}r,\\[0.4em]
&\lmatt^{(0)}(r)=\mathcal{L}_{\text{matt}0}^{(0)}
+\delta_2\frac{e^{-mr}}r
+\frac{\lmattstar^{(-2)}\delta_1}{m^2-\mu^2}
\frac{e^{-\mu r}}r,\\[0.4em]
&p(r)=p_0+\frac{2\lmattstar^{(-2)}\delta_1}{3m^2}
\frac{e^{-\mu r}}r.
\end{aligned}
\right.
\label{eq:general_solution}
\end{equation}
The background constants satisfy
\begin{align}
\mathcal{L}_{\text{matt}0}^{(0)}&=-\frac{K}{m^2},\label{eq:qn_general_backgrounds_L0} \\
p_0&=\frac{4f_0}{3g_1}-\frac{K}{3m^2}.
\label{eq:qn_general_backgrounds_p0}
\end{align}
The parameters $\delta_0,\delta_1,\delta_2,C_{\text{N}}$, and $K$ are integration constants that must be constrained by the
boundary conditions and the assumptions of the regime under consideration.

\subsection{Weak-field constraints}
\label{sec:weak_field_constraints}
To express the amplitude conditions without assigning numerical
values to the integration constants, we use the previous
solutions. Since $e^{-ar}/r$ and $(1+ar)e^{-ar}/r$ are decreasing
for $a>0$ and $r>0$, the triangle inequality yields
\begin{equation}
\begin{aligned}
G_t\leq B_t:={}&|\delta_0|+\frac{|C_\text{N}|}{r_{\min}}
+\frac{r_{\max}^2}{6|\lmattstar^{(-2)}|}
\left|\frac{f_0}{3h_{12}}+2K\right|+ \\
&\frac{4|\delta_2|e^{-mr_{\min}}}
{3|\lmattstar^{(-2)}|r_{\min}}+\\
&\frac{|\delta_1|(\mu^2+3m^2)e^{-\mu r_{\min}}}
{3\mu^2|\mu^2-m^2|r_{\min}},
\end{aligned}
\label{eq:qn_Bt_general}
\end{equation}
y
\begin{equation}
\begin{aligned}
G_r\leq B_r:={}&\frac{|C_\text{N}|}{r_{\min}}
+\frac{r_{\max}^2}{6|\lmattstar^{(-2)}|}
\left|\frac{f_0}{3h_{12}}+2K\right|+ \\
&\frac{2|\delta_2|}{3|\lmattstar^{(-2)}|}
\left(\frac1{r_{\min}}+m\right)e^{-mr_{\min}}+\\
&\frac{|\delta_1||3m^2-\mu^2|}
{3\mu^2|\mu^2-m^2|}
\left(\frac1{r_{\min}}+\mu\right)e^{-\mu r_{\min}}.
\end{aligned}
\label{eq:qn_Br_general}
\end{equation}
Here $B_t$ and $B_r$ are explicit upper bounds. The sufficient conditions
\begin{equation}
B_t\ll1,\qquad B_r\ll1,\qquad
6m^2\ell_\text{g}^2\ll1
\label{eq:qn_explicit_sufficient_conditions}
\end{equation}
guarantee a weak field on $\mathcal I$ and smallness of the
normalised metric residuals. They are sufficient, not
necessary, conditions because they do not account for possible
cancellations between the modes. In the case of identically vanishing
metric amplitudes, the corresponding normalisation is replaced by a
positive reference scale.

% ============================================================
\subsection{Asymptotically flat solutions}
\label{sec:asymptotically_flat}
% ============================================================

To select the branch of the analytical solutions that
approaches Minkowski spacetime at large radial distances,
we impose the asymptotic conditions
\cite{Wald1984}
\begin{equation}
\begin{aligned}
\lim_{r\to\infty}g_{tt}^{(2)}(r)&=0,\\
\lim_{r\to\infty}g_{rr}^{(2)}(r)&=0.
\end{aligned}
\label{eq:asymptotic_flatness}
\end{equation}
These conditions eliminate the constant contribution
to the temporal metric perturbation and the growing
terms proportional to $r^2$. Consequently,
\begin{equation}
\begin{aligned}
\delta_0&=0,\\
\frac{f_0}{3h_{12}}+2K&=0,
\end{aligned}
\label{eq:asymptotic_conditions}
\end{equation}
which gives
\begin{equation}
K=-\frac{f_0}{6h_{12}}.
\label{eq:K_asymptotic}
\end{equation}
The Newtonian normalisation of the $1/r$ contribution
allows us to identify \cite{LandauLifshitz}
\begin{equation}
C_{\mathrm{N}}=2GM,
\label{eq:CN_Newtonian}
\end{equation}
where $G$ is Newton's gravitational constant and
$M$ denotes the gravitational mass associated
with the Newtonian mode. Its physical interpretation
depends on the boundary conditions and the surrounding
matter distribution in $f(R,\lmatt)$ gravity
\cite{HarkoLobo2010}.

Using Eqs.~\eqref{eq:m2} and ~\eqref{eq:K_asymptotic},
the constant contributions to the matter variables satisfy
\begin{equation}
\begin{aligned}
\mathcal{L}_{\mathrm{matt}0}^{(0)}
&=-\frac{f_0}{g_1},\\
p_0&=\frac{f_0}{g_1},\\
\rho_0&=4\rho_*.
\end{aligned}
\label{eq:physical_backgrounds}
\end{equation}
Here, $\xi_0$, $p_0$, and $\rho_0$ denote the constant
contributions to the internal energy density,
pressure, and rest-mass energy density, respectively.
Taking into account
$\mathcal{L}_{\mathrm{matt}0}^{(0)}=\xi_0$,
we obtain
\begin{equation}
p_0=-\xi_0.
\label{eq:xi_p_background}
\end{equation}

% ============================================================
% PHYSICAL INTERPRETATION OF THE MATTER BACKGROUND
% ============================================================

\paragraph{Physical interpretation of the matter background.}

To interpret Eq.~\eqref{eq:xi_p_background}, we recall
that the Taylor expansion of $f(R,\lmatt)$ was performed
about the reference values
\begin{equation}
\begin{aligned}
R_*&=R^{(0)}=0,\\
\mathcal{L}_\text{matt*}&=\lmattstar^{(-2)}=\rho_*.
\end{aligned}
\label{eq:reference_background}
\end{equation}
Here, $R_*$ is the reference scalar curvature and
$\rho_*$ is the constant reference rest-mass energy
density. This reference configuration was chosen to
have Minkowski geometry and a dust-like material sector:
its reference pressure and internal energy density
vanish, while its matter Lagrangian is determined by
$\rho_*$. Thus, the choice of reference state is
\begin{equation}
\begin{aligned}
R_*&=0,\\
p_*&=0,\\
\xi_*&=0,\\
\mathcal{L}_\text{matt*}&= \rho_*\neq0.
\end{aligned}
\label{eq:reference_dust}
\end{equation}
The asterisk labels the reference state used to define
the Taylor coefficients, rather than a solution of the
full gravitational field equations.

We select the asymptotically flat solution branch whose
constant matter contributions retain this dust-like
character. Accordingly, we impose
\begin{equation}
\begin{aligned}
p_0&=0,\\
\xi_0&=0.
\end{aligned}
\label{eq:dust_background_condition}
\end{equation}
Combining this requirement with
Eq.~\eqref{eq:physical_backgrounds}, for $g_1\neq0$,
gives
\begin{equation}
\begin{aligned}
f_0&=0,\\
K&=0,
\end{aligned}
\label{eq:f0_dust}
\end{equation}
where $K$ is the integration constant defined in
Eq.~\eqref{eq:K_asymptotic}. The resulting constant
matter contributions are
\begin{equation}
\begin{aligned}
\rho_0&=4\rho_*\neq0,\\
p_0&=0,\\
\xi_0&=0.
\end{aligned}
\label{eq:dust_background}
\end{equation}
The constant background energy-momentum tensor
therefore has the dust form
\begin{equation}
T_{\mu\nu}^{\mathrm{bg}}=\rho_0u_\mu u_\nu,
\label{eq:dust_background_tensor}
\end{equation}
The density $\rho_0=4\rho_*$ is distinct from
the Taylor reference value $\rho_*$. The shared
dust-like character does not, by itself, establish
that the two configurations are identical or that
the Taylor truncation is valid throughout the matter
region.

The relation $p_0=-\xi_0$ should not be confused
with the vacuum-energy equation of state. Indeed,
the total energy density is
\begin{equation}
e_0=\rho_0+\xi_0,
\label{eq:background_total_energy}
\end{equation}
so that
\begin{equation}
e_0+p_0=\rho_0\neq0.
\label{eq:not_pure_vacuum}
\end{equation}
A pure vacuum-energy fluid would instead require
$p_0=-e_0$, or equivalently an energy-momentum
tensor proportional to the metric. Thus, the
selected constant matter background is dust-like,
not vacuum energy.

% ============================================================
% REDUCED ASYMPTOTICALLY FLAT SOLUTIONS
% ============================================================

\paragraph{Reduced asymptotically flat solutions.}

With $f_0=0$, the inverse radial scale $\mu$ reduces to
\begin{equation}
\mu^2=-\frac{f_1}{6f_2}>0,
\label{eq:mu_new}
\end{equation}
where $f_1$ and $f_2$ are Taylor coefficients of
the gravitational function with opposite signs.

Using $\lmatt^{(0)}=\xi$ and $\lmattstar^{(-2)}=\rho_*$,
the solutions of the reduced system become
\begin{equation}
\left\{
\begin{aligned}
&g_{tt}^{(2)}(r)
=-\frac{2GM}{r}
-\frac{4\delta_2}{3\rho_*}
\frac{e^{-mr}}{r}
+\frac{\delta_1(\mu^2+3m^2)}
{3\mu^2(\mu^2-m^2)}
\frac{e^{-\mu r}}{r},
\\[0.8em]
&\begin{aligned}
g_{rr}^{(2)}(r)
=&-\frac{2GM}{r}
-\frac{2\delta_2}{3\rho_*}
(1+mr)\frac{e^{-mr}}{r}+\\
&\frac{\delta_1(3m^2-\mu^2)}
{3\mu^2(\mu^2-m^2)}
(1+\mu r)\frac{e^{-\mu r}}{r},
\end{aligned}
\\[0.8em]
&R^{(2)}(r)
=\delta_1\frac{e^{-\mu r}}{r},
\\[0.8em]
&\rho(r)
=4\rho_*+
A\frac{e^{-mr}}{r},
\\[0.8em]
&\xi(r)
=\delta_2\frac{e^{-mr}}{r}+\frac{\rho_*\delta_1}{m^2-\mu^2}
\frac{e^{-\mu r}}{r},
\\[0.8em]
&p(r)
=\frac{2\rho_*\delta_1}{3m^2}
\frac{e^{-\mu r}}{r}.
\end{aligned}
\right.
\label{eq:physical_solution}
\end{equation}
For $m>0$ and $\mu>0$, the formal radial continuation
satisfies
\begin{equation}
\begin{aligned}
\lim_{r\to\infty}p(r)&=0,\\
\lim_{r\to\infty}\xi(r)&=0,\\
\lim_{r\to\infty}\rho(r)&=4\rho_*.
\end{aligned}
\label{eq:dust_formal_limit}
\end{equation}
These limits describe the analytical continuation of
the matter profiles and do not require the physical
material distribution to extend to infinity.

For a finite matter region, we consider the interval $\mathcal I=[r_{\min},r_{\max}]$.
Within this interval, an approximately dust-like
regime may occur when $\rho(r)>0$ and
\begin{equation}
\begin{aligned}
|p(r)|&\ll\rho(r),\\
|\xi(r)|&\ll\rho(r).
\end{aligned}
\label{eq:approximately_dust}
\end{equation}
A globally asymptotically flat configuration with
compactly supported matter would additionally
require matching to an exterior vacuum solution
under the junction conditions of the theory.

% ============================================================
% WEAK-FIELD BOUNDS
% ============================================================

For this asymptotically flat branch, sufficient
weak-field bounds on the finite interval can be
written as
\begin{equation}
\begin{aligned}
B_t={}&
\frac{2G|M|}{r_{\min}}
+\frac{4|\delta_2|e^{-mr_{\min}}}
{3|\lmattstar^{(-2)}|r_{\min}}
+\frac{|\delta_1|(\mu^2+3m^2)e^{-\mu r_{\min}}}
{3\mu^2|\mu^2-m^2|r_{\min}},
\end{aligned}
\label{eq:qn_Bt_flat}
\end{equation}
and
\begin{equation}
\begin{aligned}
B_r={}&
\frac{2G|M|}{r_{\min}}
+\frac{2|\delta_2|}
{3|\lmattstar^{(-2)}|}
\left(\frac{1}{r_{\min}}+m\right)e^{-mr_{\min}}+\\
&\frac{|\delta_1||3m^2-\mu^2|}
{3\mu^2|\mu^2-m^2|}
\left(\frac{1}{r_{\min}}+\mu\right)e^{-\mu r_{\min}}.
\end{aligned}
\label{eq:qn_Br_flat}
\end{equation}
Although the constant matter background is dust-like,
the complete matter distribution generally has
nonzero pressure and internal energy density. It is
therefore described by the perfect-fluid tensor
\eqref{eq:T_mu_nu}, rather than by pressureless dust.

A particular branch is obtained by setting $\delta_2=0$.
In this case, the internal energy density and pressure
reduce to
\begin{equation}
\begin{aligned}
\xi(r)
&=\frac{\rho_*\delta_1}{m^2-\mu^2}
\frac{e^{-\mu r}}{r},\\
p(r)
&=\frac{2\rho_*\delta_1}{3m^2}
\frac{e^{-\mu r}}{r}.
\end{aligned}
\label{eq:gamma_law_profiles}
\end{equation}
For $\delta_1\neq0$, the ratio between pressure and
internal energy density is independent of the radial
coordinate,
\begin{equation}
\frac{p(r)}{\xi(r)}
=\frac{2(m^2-\mu^2)}{3m^2}.
\label{eq:pressure_internal_energy_ratio}
\end{equation}
Consequently, this branch satisfies the relation
\begin{equation}
p(r)=(\kappa-1)\xi(r),
\label{eq:gamma_law}
\end{equation}
where the constant parameter $\kappa$ is
\begin{equation}
\kappa=\frac{5}{3}-\frac{2\mu^2}{3m^2}.
\label{eq:kappa_expression}
\end{equation}
For $0<\mu^2<m^2$, one obtains
$1<\kappa<5/3$. Equation~\eqref{eq:gamma_law}
coincides with the internal-energy relation associated
with the relativistic polytropic equation of state
considered by Tooper~\cite{Tooper1965}.

We now examine whether the matter profiles can also
satisfy a polytropic relation of the form
\begin{equation}
p(r)=K_{\mathrm{poly}}\rho(r)^\kappa,
\label{eq:polytropic_eos}
\end{equation}
where $K_{\mathrm{poly}}$ is a constant.
Rather than imposing this relation globally, we
consider its local validity around a fixed reference
radius $R>0$, within the radial domain of the
weak-field solution.

Evaluating Eq.~\eqref{eq:polytropic_eos} at $r=R$
gives
\begin{equation}
K_{\mathrm{poly}}
=
\frac{2\rho_*\delta_1 e^{-\mu R}}
{3m^2 R
\left(4\rho_*+Ae^{-mR}/R\right)^\kappa}.
\label{eq:polytropic_constant}
\end{equation}
To require agreement also to first order around
$R$, we impose
\begin{equation}
\left.
\frac{p'(r)}{p(r)}
\right|_{r=R}
=
\kappa
\left.
\frac{\rho'(r)}{\rho(r)}
\right|_{r=R}.
\label{eq:polytropic_local_condition}
\end{equation}
Substituting the analytical profiles yields
\begin{equation}
(\mu R+1)
\left(4\rho_*R+Ae^{-mR}\right)
=
\kappa Ae^{-mR}(mR+1).
\label{eq:polytropic_matching}
\end{equation}
Solving for the integration constant $A$, we obtain
\begin{equation}
A=
\frac{4\rho_*R(\mu R+1)e^{mR}}
{\kappa(mR+1)-(\mu R+1)}.
\label{eq:A_polytropic}
\end{equation}
For $0<\mu<m$ and any finite reference radius $R>0$, the denominator of Eq.~\eqref{eq:A_polytropic} is strictly positive, since $\kappa-1>0$ and $\kappa m-\mu>0$. Consequently, the integration constant $A$ is finite for finite $\rho_*$.

Using Eq.~\eqref{eq:A_polytropic}, the density
profile becomes
\begin{equation}
\rho(r)=4\rho_*\left[1 +
\frac{R(\mu R+1)}
{\kappa(mR+1)-(\mu R+1)}
\frac{e^{-m(r-R)}}{r}\right].
\label{eq:rho_local_polytropic}
\end{equation}
The corresponding polytropic constant is
\begin{equation}
\begin{aligned}
K_{\mathrm{poly}}
={}&
\frac{2\rho_*\delta_1e^{-\mu R}}
{3m^2R}
\left[
\frac{\kappa(mR+1)-(\mu R+1)}
{4\rho_*\kappa(mR+1)}
\right]^\kappa.
\end{aligned}
\label{eq:K_polytropic_final}
\end{equation}
The resulting profiles satisfy the polytropic
relation and its first radial derivative at $r=R$.
Equivalently, defining
\begin{equation}
\mathcal{Q}(r):=
\frac{p(r)}
{K_{\mathrm{poly}}\rho(r)^\kappa},
\label{eq:polytropic_ratio}
\end{equation}
we obtain
\begin{equation}
\mathcal{Q}(R)=1,
\qquad
\mathcal{Q}'(R)=0.
\label{eq:polytropic_local_agreement}
\end{equation}
Therefore, for sufficiently small $|r-R|$,
\begin{equation}
\mathcal{Q}(r)
=
1+\mathcal{O}\!\left((r-R)^2\right).
\label{eq:polytropic_local_expansion}
\end{equation}
This establishes a local first-order polytropic
approximation, rather than an exact global
polytropic equation of state. Its physical
applicability remains subject to the weak-field
and matter-perturbation conditions adopted
in the present work.

Under these conditions, the analytical solutions corresponding to the locally polytropic branch can be expressed as
\begin{equation}
\left\{
\begin{aligned}
&g_{tt}^{(2)}(r)
=-\frac{2GM}{r}
+\frac{\delta_1(\mu^2+3m^2)}
{3\mu^2(\mu^2-m^2)}
\frac{e^{-\mu r}}{r},
\\[0.8em]
&\begin{aligned}
g_{rr}^{(2)}(r)
=&-\frac{2GM}{r}
+\frac{\delta_1(3m^2-\mu^2)}
{3\mu^2(\mu^2-m^2)}
(1+\mu r)\frac{e^{-\mu r}}{r},
\end{aligned}
\\[0.8em]
&R^{(2)}(r)
=\delta_1\frac{e^{-\mu r}}{r},
\\[0.8em]
&\rho(r)=4\rho_*\left[1 +
\frac{R(\mu R+1)}
{\kappa(mR+1)-(\mu R+1)}
\frac{e^{-m(r-R)}}{r}\right],
\\[0.8em]
&\xi(r)
=\frac{\rho_*\delta_1}{m^2-\mu^2}
\frac{e^{-\mu r}}{r},
\\[0.8em]
&p(r)
=\frac{2\rho_*\delta_1}{3m^2}
\frac{e^{-\mu r}}{r}.
\end{aligned}
\right.
\label{eq:polytropic_solutions}
\end{equation}
Here, $\delta_1$ determines the amplitude of the Yukawa-type curvature
perturbation and its associated contributions to the
metric, internal energy density, and pressure.
Its value and sign are constrained by the physical
conditions imposed on the solutions.

\section{Spacetime and extended Newtonian potential}

Using the metric perturbations obtained for the locally
polytropic branch in Eq.~\eqref{eq:polytropic_solutions},
the corresponding spacetime geometry in the weak-field
regime is described by the line element
\begin{equation}
\begin{aligned}
\text{d}s^2=&\Bigg[
1-\frac{2GM}{r}
+
\frac{\delta_1(\mu^2+3m^2)}
{3\mu^2(\mu^2-m^2)}
\frac{e^{-\mu r}}{r}
\Bigg]\text{d}t^2-\\[1ex]
&\Bigg[
1+\frac{2GM}{r}
-
\frac{\delta_1(3m^2-\mu^2)}
{3\mu^2(\mu^2-m^2)}
(1+\mu r)\frac{e^{-\mu r}}{r}
\Bigg]\text{d}r^2
-\\ &r^2\text{d}\Omega^2 .
\end{aligned}
\label{eq:full_metric}
\end{equation}
This metric describes the gravitational field of a
compact object surrounded by the matter distribution
associated with the locally polytropic branch of
$f(R,\lmatt)$ gravity, within the weak-field regime
considered in this work.

\subsection{Extended Newtonian potential}
\label{sec:extended_newtonian_potential}

In the weak-field regime, the perturbation of the temporal
component of the metric is related to the gravitational
potential through~\cite{CapozzielloDeLaurentis2012}
\begin{equation}
g_{tt}^{(2)}(r)=2\Phi_{\text{grav}}(r).
\label{eq:gtt_potential}
\end{equation}
Using the solution \eqref{eq:polytropic_solutions},
the gravitational potential can be written as
\begin{equation}
\Phi_{\text{grav}}(r)
=
-\frac{GM}{r}
\left(
1+\alpha_\mu e^{-\mu r}
\right),
\label{eq:newtonian_potential}
\end{equation}
where
\begin{equation}
\alpha_\mu
:=
\frac{\delta_1(\mu^2+3m^2)}
{6GM\mu^2(m^2-\mu^2)}.
\label{eq:alpha_mu}
\end{equation}
The first term reproduces the Newtonian contribution
associated with the $1/r$ mode, while the second
introduces a Yukawa-type correction characterised by
the inverse length scale $\mu$. This type of correction
appears naturally in the weak-field limit of higher-order
gravitational theories
~\cite{Stelle1978,CapozzielloStabileTroisi2009}.

The radial gravitational acceleration associated with
the potential \eqref{eq:newtonian_potential} is
\begin{equation}
a_r(r)
=
-\frac{\text{d}\Phi_{\text{grav}}}{\text{d}r}
=
-\frac{GM}{r^2}
\left[
1+\alpha_\mu(1+\mu r)e^{-\mu r}
\right].
\label{eq:radial_acceleration}
\end{equation}
For $GM>0$, the gravitational interaction is attractive
when
\begin{equation}
1+\alpha_\mu(1+\mu r)e^{-\mu r}>0,
\label{eq:attractive_condition}
\end{equation}
whereas a repulsive radial acceleration occurs when
\begin{equation}
1+\alpha_\mu(1+\mu r)e^{-\mu r}<0.
\label{eq:repulsive_condition}
\end{equation}
Since $\mu>0$, the function
$(1+\mu r)e^{-\mu r}$ is positive and monotonically
decreasing for $r>0$, approaching unity as $r\to0^+$
and vanishing as $r\to\infty$. Consequently, the
gravitational acceleration remains attractive for
$\alpha_\mu\geq-1$, whereas $\alpha_\mu<-1$
allows a repulsive region at sufficiently small radii.

Furthermore, under the condition $0<\mu<m$,
the sign of $\alpha_\mu$ is determined by the
integration constant $\delta_1$. In particular,
$\delta_1>0$ corresponds to an attractive Yukawa
correction, while $\delta_1<0$ allows a repulsive
contribution, which can dominate the Newtonian
term only when $\alpha_\mu<-1$.

The existence of such a repulsive region is a
mathematical possibility of the effective potential.
Its physical applicability requires that the
QN weak-field limit and matter-perturbation
conditions remain satisfied throughout the
corresponding radial domain.

The Yukawa correction is exponentially suppressed
at large distances, i.e. when:
\begin{equation}
\Phi_{\text{grav}}(r)
\sim -\frac{GM}{r},
\qquad r\to\infty,
\label{eq:newtonian_asymptotic}
\end{equation}
so that the Newtonian behaviour is recovered
asymptotically. The quantity $\mu^{-1}$ determines
the characteristic spatial scale over which the
Yukawa correction can be relevant.

\section{Conclusions}
\label{sec:conclusions}

In this work, we have investigated the
weak-field regime of $f(R,\lmatt)$ gravity
for a static, spherically symmetric
configuration containing a central compact
object surrounded by a perfect fluid.
By performing a post-Newtonian expansion
of the gravitational field equations and
introducing a quasi-Newtonian approximation,
we obtained analytical solutions for the
metric perturbations, the Ricci scalar,
and the matter variables.

The general analytical solutions contain
the usual Newtonian contribution together
with two Yukawa-type corrections,
characterised by the length scales
$m^{-1}$ and $\mu^{-1}$.
These corrections arise from the coupling
between curvature and matter and introduce
scale-dependent modifications to the
weak-field gravitational field.

We also identified a particular solution
branch, obtained by setting $\delta_2=0$,
in which the pressure and internal energy
density satisfy a $\kappa$-law relation.
Within this branch, a local first-order
polytropic approximation can be established
at a finite reference radius.
The corresponding spacetime metric contains
a single Yukawa correction characterised
by the inverse length scale $\mu$.

The extended Newtonian potential associated
with this branch admits both attractive
and repulsive Yukawa contributions,
depending on the integration constant
$\delta_1$.
In particular, a repulsive radial
acceleration is mathematically possible
when the Yukawa contribution dominates
the Newtonian term.
The physical realisation of this behaviour
remains subject to the validity of the
approximations adopted in this work.

The analytical solutions are applicable
within the weak-field and quasi-Newtonian
regimes established here.
In particular, the quasi-Newtonian bounds
control the omitted terms relative to the
adopted differential scales, without
establishing the validity of the approximation
beyond the specified radial domain.

These results provide an analytical
framework for investigating the effects
of curvature--matter couplings on the
gravitational field of compact objects
surrounded by matter and motivate further
studies of their physical implications.

\section{Acknowledgements}
This work was supported by a PAPIIT DGAPA-UNAM grant IN118325. SM and RU acknowledge support from Secihti (26344, 1264063).

%================================================
% Bibliography
%================================================

\end{document}